\documentclass[aps,amssymb,amsmath, twocolumn, pra, superscriptaddress]{revtex4}
\usepackage{graphicx}
\usepackage{epsfig}
\usepackage{dcolumn}
\usepackage[up]{subfigure}
\usepackage{float}
\usepackage{dsfont}	
\usepackage{relsize}

\usepackage{epstopdf}

\usepackage{float}
\usepackage{ragged2e}
\usepackage{braket}
\usepackage[font=small,labelfont=bf,justification=raggedright, format=plain]{caption}
\usepackage{setspace}
\usepackage[colorlinks, citecolor = magenta]{hyperref}
\makeatletter

\begin{document}

\title{Polarization-path-frequency entanglement using interferometry and frequency shifters}


\author{Mrittunjoy Guha Majumdar}
\email{mrittunjoyg@iisc.ac.in}
\affiliation{Quantum Optics \& Quantum Information, Department of Instrumentation and Applied Physics, Indian Institute of Science, Bengaluru, India}
\author{C. M. Chandrashekar}%
\email{chandracm@iisc.ac.in}
\affiliation{Quantum Optics \& Quantum Information, Department of Instrumentation and Applied Physics, Indian Institute of Science, Bengaluru, India}
\affiliation{The Institute of Mathematical Sciences, C. I. T. Campus, Taramani, Chennai 600113, India}
\affiliation{Homi Bhabha National Institute, Training School Complex, Anushakti Nagar, Mumbai 400094, India}


\begin{abstract}
Higher dimensional Hilbert space along with ability to control  multiple degrees of freedom of photon and entangle them has enabled new quantum protocols for various quantum information processing applications. Here, we propose a  scheme to generate  and control polarization-path-frequency entanglement using the operative elements required to implement a polarization-controlled quantum walk in the path(position) space and frequency domain.  Hyperentangled states manifests in the controlled dynamics using an interferometric setup where  half-wave plates, beam-splitters and frequency shifters such as those based on the electro-optic effect are used to manipulate the polarization, path and frequency degrees of freedom respectively. The emphasis is on utilizing the polarization to influence the movement to a specific value in the frequency and position space.  Negativity between the subspaces is calculated to demonstrate the controllability of the entanglement between them.  Progress reported with experimental demonstration of   realization of quantum walk using quantum states of light makes quantum walks a practical approach to generate hyperentangled states. 
\end{abstract}

\maketitle


 \section{Introduction}

 Entanglement has been the underlying resource for various quantum mechanical applications. Entanglement across and within varied degrees of freedom quantum particle has been even more useful in optimizing the number of physical qubits and the logical quantum mechanical operations and applications one can undertake with it\,\cite{horodecki2009quantum, plenio2014introduction, weiner2002entanglement, deng2017quantum, xie2015harnessing}. When it comes to photonic qubits, we have had hyperentanglement across various degrees of freedom: polarization-path\,\cite{ciampini2016path}, polarization-OAM\,\cite{majumdar2020quantum}, polarization-frequency\,\cite{shi2012compact, prabhu2021hyperentanglement} as well as polarization-time bin\,\cite{du2016general, chapman2019hyperentangled}. The level of advancement in the realm of hyperentanglement has reached a point where we can now simultaneously use three degrees of freedom for quantum information processing\,\cite{wang201818, xu2012proposal}. Applications of hyperentanglement have included entanglement purification and hyperdistillation\,\cite{zhou2021high, wang2015one}, Bell state and GHZ state analysis\,\cite{walborn2003hyperentanglement, song2013complete}, high-capacity quantum secure direct communication\,\cite{tie2011high}, quantum error correction\,\cite{wilde2009linear}, quantum superdense coding\,\cite{rui2012quantum}, quantum secret sharing\,\cite{gu2012high} and quantum sensing\,\cite{smith2015sensors}.  With continued progress being reported establishing the control over various combination of degrees of freedom of light,  combining them by simultaneously controlling  more than three degree of freedom for quantum information processing task can be envisioned.  
 
 A quantum walk is the quantum version of the classical random walk, and in the form of algorithm it has shown significant advantages over classical algorithms\,\cite{childs2002example}. The advantage of a quantum walk comes from the way it evolves the walker in a superposition of position space states entangling the Hilbert spaces associated with the walker. Such quantum walks have been ubiquitous in various applications such as search algorithms\,\cite{travaglione2002implementing, shenvi2003quantum}. While Richard Feynman proposed the possibility of implementation of universal quantum computation with a time-dependent Hamiltonian\,\cite{feynman1985quantum}, its possibility was shown when the Hamiltonian is limited to being the adjacency matrix of a low-degree graph\,\cite{childs2009universal}. The resultant continuous-time quantum walk was found to be a universal quantum computation primitive.
Recent works have shown that the discrete-time quantum walks are also universal for quantum computation\,\cite{lovett2010universal, singh2021universal} as are multi-particle quantum walks in interacting systems such as the Bose-Hubbard model\,\cite{childs2013universal}. Quantum walks has also been one of the effective algorithmic approach to generate controlled dynamics and engineer entanglement between subsystems that are part of the dynamics\,\cite{chandrashekar2012disorder}. Recently, 
hyperentangled states generated in discrete-time quantum walks have been used as a resource
for efficacious and viable quantum dialogue protocol\,\cite{liu2020quantum}. New schemes for realization of a quantum walk based hyperentangled states of photons using J - (q)- plates and polarization beamsplitters has also been proposed\,\cite{yasir2021generation}. 

To use frequency of photons as effective degree of freedom to generate higher dimensional entangled states, conversion of frequency and ability to retain them in superposition of different frequencies are important. The optical frequency bands for storage and transmission of optical photon qubits are distinct. The inter-convertibility between the two regimes is pivotal to a seamless integration of quantum computing and communication technologies. A recent experiment demonstrates the coherent frequency conversion of
photons entangled in their polarization\,\cite{ramelow2012polarization}. This has naturally led to the prospect of changing the photon's spectral bandwidth during frequency conversion by suitably designing phase matching. But what is even more interesting is the possibility of accessing myriad states in the higher-dimensional Hilbert Space of the frequency degree-of-freedom. The addressability of these states can be made precise by specific hyperentangled structures that establish a one-to-one correspondence between a higher dimensional space qubit and an associated hyperentangled qubit that tags it. In our scheme, we look at doing this selectively by applying an (hyper-conditional) operation depending on what the value of the `tagging' degree-of-freedom is, in an interferometric setup. This augments the higher-dimensional nature of hyperentangled states by harnessing the resources of multiple degrees-of-freedom as well as the multiplicity inherent in single, specific degrees-of-freedom.    

Higher-dimensional Hilbert Spaces, particularly when tagged by one or more associated hyperentangled degrees-of-freedom, can be effective in facilitating quantum information processing tasks. Quantum hyperdense coding protocol in higher dimensions was utilized for a quantum secure direct communication (QSDC) protocol and was shown to have a high capacity and better security than that we can obtain with a two-dimensional entangled quantum system\,\cite{wang2005quantum}. A platform for photon-efficient quantum communication with energy–time–polarization high-dimensional encoding has been proposed using a biphoton frequency comb technique\,\cite{xie2015harnessing}. Recently a higher-dimensional measurement-device-independent quantum key distribution (MDI-QKD) protocol in multiple DOFs which is based on a linear-optical hyperentangled Bell state analysis has also been proposed\,\cite{yang2021feasible}. Most applications of the frequency degree-of-freedom in a higher-dimensional entangled system are premised on distinct frequency shifts and usually on two frequency states, such as in the case of remote state preparation of single-photon states in polarization and frequency DoFs\,\cite{wang2021remote}. While there has been discussion on spectral linear optical quantum computation\,\cite{lukens2017frequency}, the spectrum of frequency modes have not been utilized in quantum information processing, particularly with discernment of individual modes over a large spread of frequencies.

In this work, we propose a scheme for harnessing path-polarization-frequency hyperentanglement by undertaking a polarization-controlled quantum walk in the frequency domain and position space. We would like to make a note that the path and position degree of freedom are treated equally in this work. Our proposal has an interferometry approach where polarizing beam splitters and frequency shifter are used with a control on polarization state using a half-wave plate to control. Frequency shifters are like those that use the electro-optic effect\,\cite{minet2020pockels, preble2007changing, wright2017spectral, hu2021chip} to shift the frequency of light using variations in the birefringence in an optical medium induced by an electric field.  We also show the way entanglement between the sub-spaces manifests with the dynamics.

In section\,\ref{model}, we discuss discrete-time quantum walks and present our model for polarization-dependent discrete-time quantum walk in the position and frequency space and the way hyperentangled states manifests in the dynamics. In the same section we also discuss the current state of art in resolving the frequency shifts generated by frequency shifters like electro-optic modulators. In section\,\ref{conc} we conclude with our remarks.

\section{The Model}
\label{model}
The quantum walk is distinct from its classical counterpart, particularly in that there is a ballistic spread of the probability distribution of the walker due to interference effects, unlike when there is a diffusive effect as in the case of the classical walk. Quantum walks are studied in both, continuous-time and discrete time frame work. In this work we will focus on discrete-time version of the quantum walk\,\cite{meyer1996quantum, ambainis2001one, chanSu2}. In a discrete-time quantum walk, the walker is a quantum system which resides on the composite Hilbert space $\mathcal{H}_{T} = \mathcal{H}_c \otimes \mathcal{H}_p$. An internal degree of freedom of the walker spans the basis state of the coin Hilbert space $\mathcal{H}_c$ and the position space on which the walker spreads spans the basis state of the  position Hilbert space $\mathcal{H}_p$. For a walker on a one-dimensional position space the basis state of $\mathcal{H}_c$ will be $|0\rangle$ and $|1\rangle$ and the basis state of the $\mathcal{H}_p$ will be $\{\ket{x}\}$ where $x  \in \mathbb{Z}$ represents the labels on the position space. The general form of  initial state of the walker can be written as, 
\begin{equation}
\ket{\Psi}_{in} = \cos(\delta) \ket{0} + e^{-i\eta} \sin(\delta) \ket{1} \otimes \ket{x =0}.
\end{equation}
Each step of the walk evolution in one-dimensional position space is defined using a quantum coin operation acting on the coin space 
\begin{equation}
\mathcal{C}(\theta) = \begin{bmatrix}
   \cos(\theta) & - \sin(\theta) \\
   \sin(\theta) & ~~ \cos(\theta) 
  \end{bmatrix}
  \label{qcoin}
\end{equation}
followed by a conditional position shift operation acting on the full Hilbert space
\begin{equation} \label{Shift}
\mathcal{S}  = \sum_{x\in\mathbb{Z}} \bigg (\ket{0}\bra{0}
\otimes   \ket{x-1}\bra{x}+\ket{1}\bra{1} \otimes \ket{x+1}\bra{x}\bigg ).
\end{equation}
The state after $t$ steps of the walk will be 
\begin{equation}
\ket{\Psi}_t = \Big[\mathcal{S} (\mathcal{C}(\theta) \otimes  \mathbb{I}_{x})\Big ]^t \ket{\Psi}_{in}.
\end{equation}
By using different variant of the coin and shift operators, different forms of discrete-time quantum walks can be realized\,\cite{chandrashekar2012disorder, SCB21}. 

To generate hyperentanglement in polarization-frequency-position space of the photon we extend the above described evolution to the higher dimensional Hilbert space $\mathcal{H}_{Hyp} = \mathcal{H}_c \otimes \mathcal{H}_p \otimes \mathcal{H}_f$ associated with the photons. The $\mathcal{H}_c$ will span the polorization degree of freedom $\ket{H}$ and $\ket{V}$, the $\mathcal{H}_p$ will span the position space $\ket{x_i}$ and the $\mathcal{H}_f$ will span the frequency space $\ket{f_i}$. The operations to define the walk on these three degree of freedom will be a composition of the coin operation on the polarization space followed by a polarization conditioned shift operation on the position and frequency space. The Half-Wave plate (HWP) executes the rotation $C(\theta)$ given by the Eq.\,\eqref{qcoin} on the polarization space. The polarization Beamsplitter (PBS) executes a polarization conditioned shift operation of the form
\begin{align}
\label{ShiftP}
\mathcal{S_P}  = & \Bigg [ \ket{H}\bra{H} \otimes \ket{x+1}\bra{x} \nonumber \\
& + \sum_{i\, \in \, \mathbb{Z}} \bigg (\ket{V}\bra{V}
\otimes   \ket{x-1}\bra{x} \bigg ) \Bigg ] \otimes  \mathbb{I}_{f}.
\end{align} 
Electro-optics frequency shifters (EOFSs) are not polarization dependent shifters and therefore cannot be directly used as a polarization conditioned shift operator as in the form of Eq.\,\eqref{ShiftP}. The electro-optic effect mixes optical and microwave fields directly, and in recent on-chip realisations of frequency shifting, the frequency of all inserted photons were seen to be shifted to another frequency in the regime of full conversion \cite{hu2021chip}. This allows us tune a single frequency shift and make it evolve in superposition with the polarization state of the photon. A polarization dependent shift operator using EOMs can be executed by placing the EOM (or any frequency shifter) in one of the path after the two polarization states are spatially separated. The operational representation will be of the form, 
\begin{align}
\label{Shiftf}
\mathcal{S}_f  = & \Bigg [ \ket{H}\bra{H} \otimes  \mathbb{I}_{x} \otimes  \ket{f_{i+1}}\bra{f_i}  \nonumber \\
& + \sum_{i\, \in \, \mathbb{Z_+}} \bigg (\ket{V}\bra{V}
\otimes    \mathbb{I}_{x} \otimes \ket{f_i}\bra{f_i}\bigg ) \Bigg ] .
\end{align} 
\begin{figure}[ht!]
   \centering
    \includegraphics[width=0.47\textwidth]{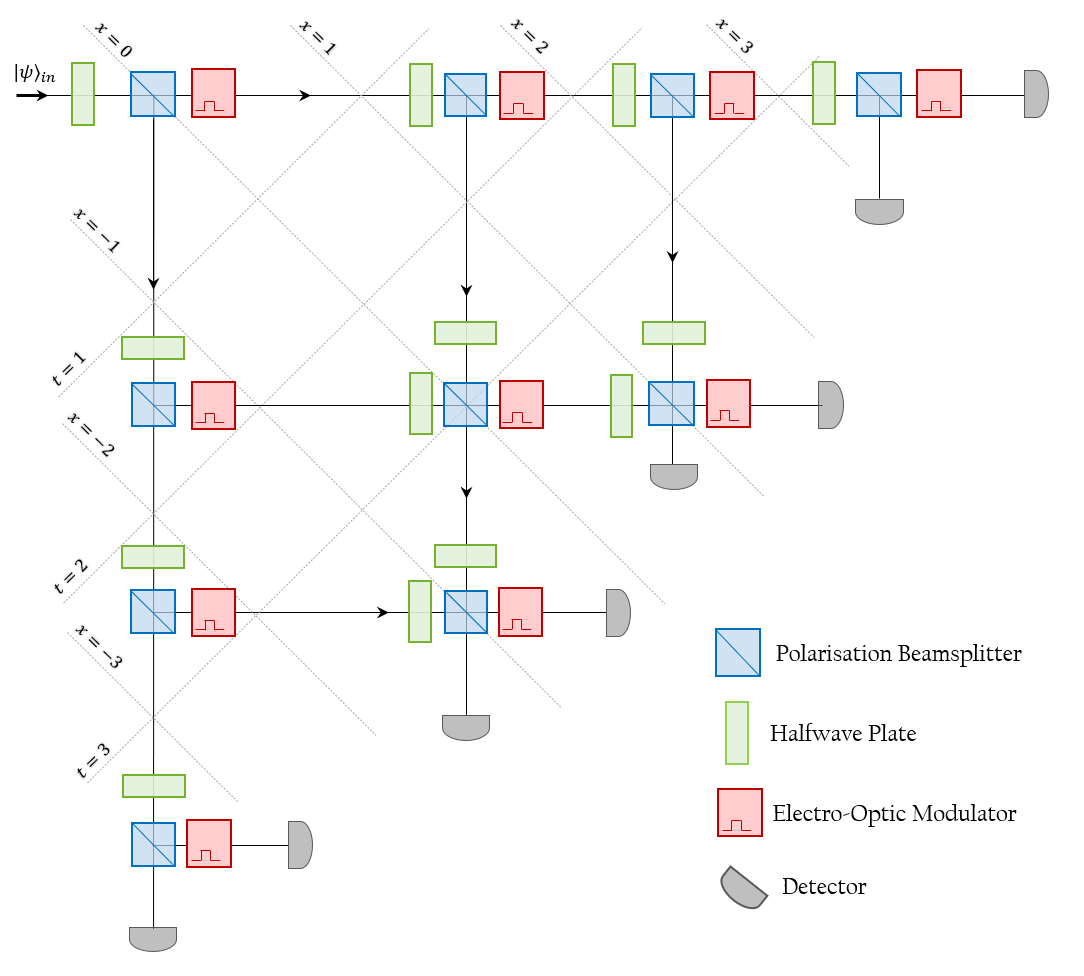}
    \caption{Schematic of optical elements with spatial and temporal markings to realize polarization-controlled quantum walk in the frequency and position degrees-of -freedom. Each step of walk is a composition of two HWP, two PBS and one EOM. EOMs are placed across the path where only state $\ket{H}$ pass through. Each detector will capture different frequencies with the difference between them being a function of the electro-optic frequency shifts carried out by the EOMs.}
   \label{fig:my_label}
\end{figure}
This form of shift operators has been effectively used in defining directed quantum walks\,\cite{MayDiretQW, ChaPerco}.  
Thus, for a photon in initial state
\begin{align}
|\Psi_p\rangle_{in} = \bigg (\cos(\delta) |H\rangle - \sin(\delta) \ket{V} \bigg ) \otimes \ket{x_0}\bra{x_0} \otimes \ket{f_0} \bra{f_0}
\end{align}
each step of walk operation will be a composition of evolution in superposition of position space followed by the evolution in superposition of frequency space,
\begin{align}
\mathcal {W} =  \mathcal{S}_f \mathcal{S_P} \bigg ( \mathcal{C}(\theta) \otimes  \mathbb{I}_{x}\otimes  \mathbb{I}_{f_i} \bigg). 
\end{align}
In Fig.\,\ref{fig:my_label}  we give a schematic of the optical setup for four step of discrete-time quantum walk composing of HWPs, PBSs, and EOMs to generate polarization-path-frequency  entanglement. Two HWP, two PBS and one EOM is required for each step of the walk. As shown in the schematic the detector are placed at the end identified with the position space and each position space will uniquely capture different frequencies with the difference between the adjacent detectors will be the frequency of the acoustic wave used to drive frequency shift. After $t$ steps of walk evolution in network of optical setup will be
\begin{align}
|\Psi_p(t)\rangle & =  \mathcal{W}^t  |\Psi_p\rangle_{in}  \nonumber \\
 &  =\sum_{k=0}^t \Bigg [ \xi^H_{k}(\delta, \theta)\vert H\rangle \vert f_{t} \rangle  \vert x_{-n+2t}\rangle \nonumber \\
 & ~~~~~~~~~~~+ \xi^V_{k}(\delta, \theta) \vert V\rangle \vert f_{t}\rangle \vert x_{-n+2t}\rangle \Bigg ].
\end{align}
The above state is a Schmidt rank vector of $(2,t,t)$.  The probability distribution in position space will be 
\begin{align}
P(x, t) = \sum_x \bra{x} \bigg ( \ket{\Psi_p (t)} \bra{\Psi_p(t)} \bigg ) \ket{x}
\end{align}
and in frequency space it will be 
\begin{align}
P(f_i, t) = \sum_i \bra{f_i} \bigg ( \ket{\Psi_p (t)} \bra{\Psi_p(t)} \bigg ) \ket{f_i}.
\end{align}

\begin{figure}[h!]
   \centering
    \includegraphics[width=0.47\textwidth]{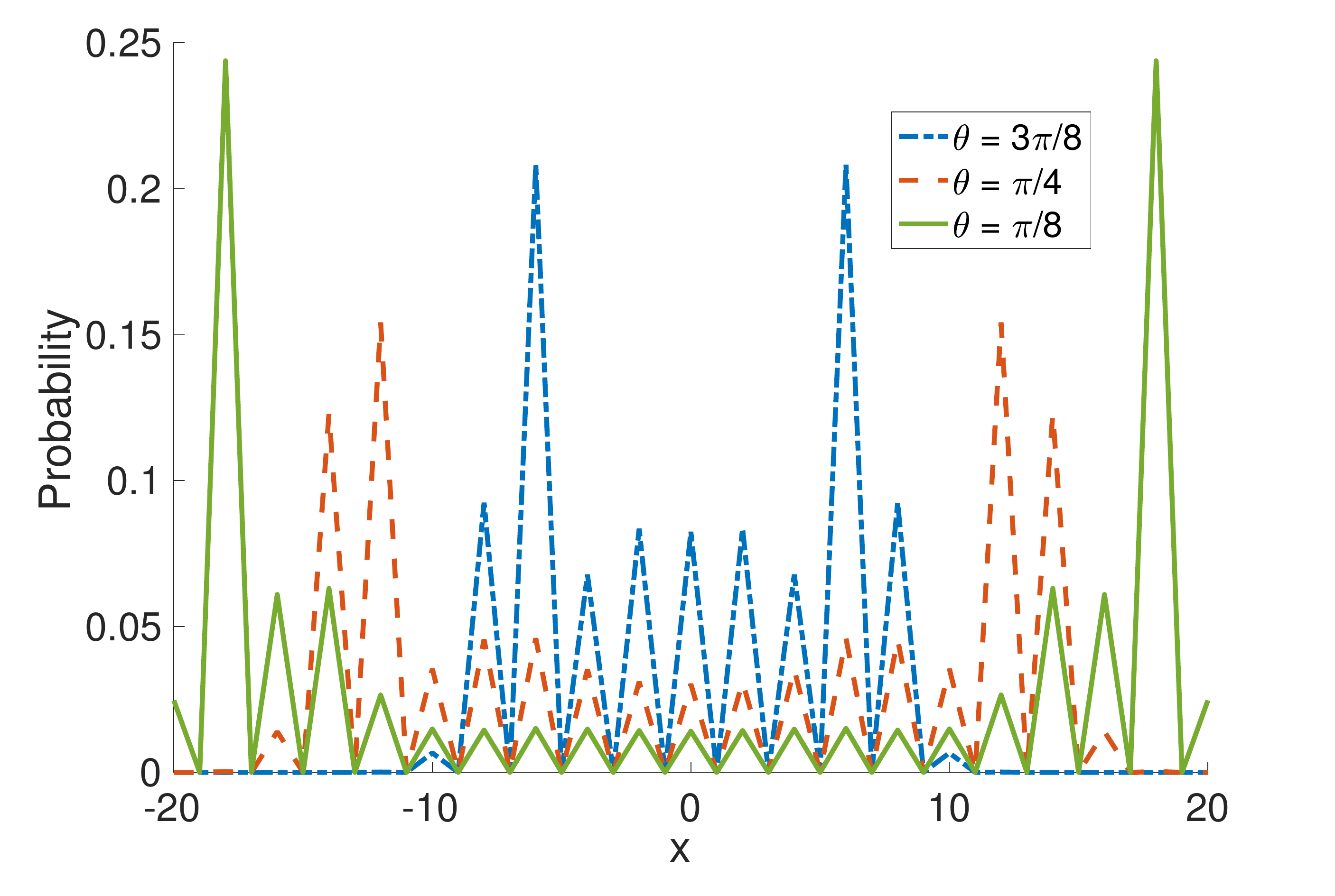}
    \caption{Probability distribution in position space after 20 step of walk using different rotation angles $\theta$. We can see that the probability of finding the states is zero for alternate position space. When this is compared with the schematic in Fig.\,\ref{fig:my_label}, detectors are places only at odd(even) positions after odd (even) number of steps.}
   \label{fig:prob1}
\end{figure}

\begin{figure}[h!]
   \centering
    \includegraphics[width=0.47\textwidth]{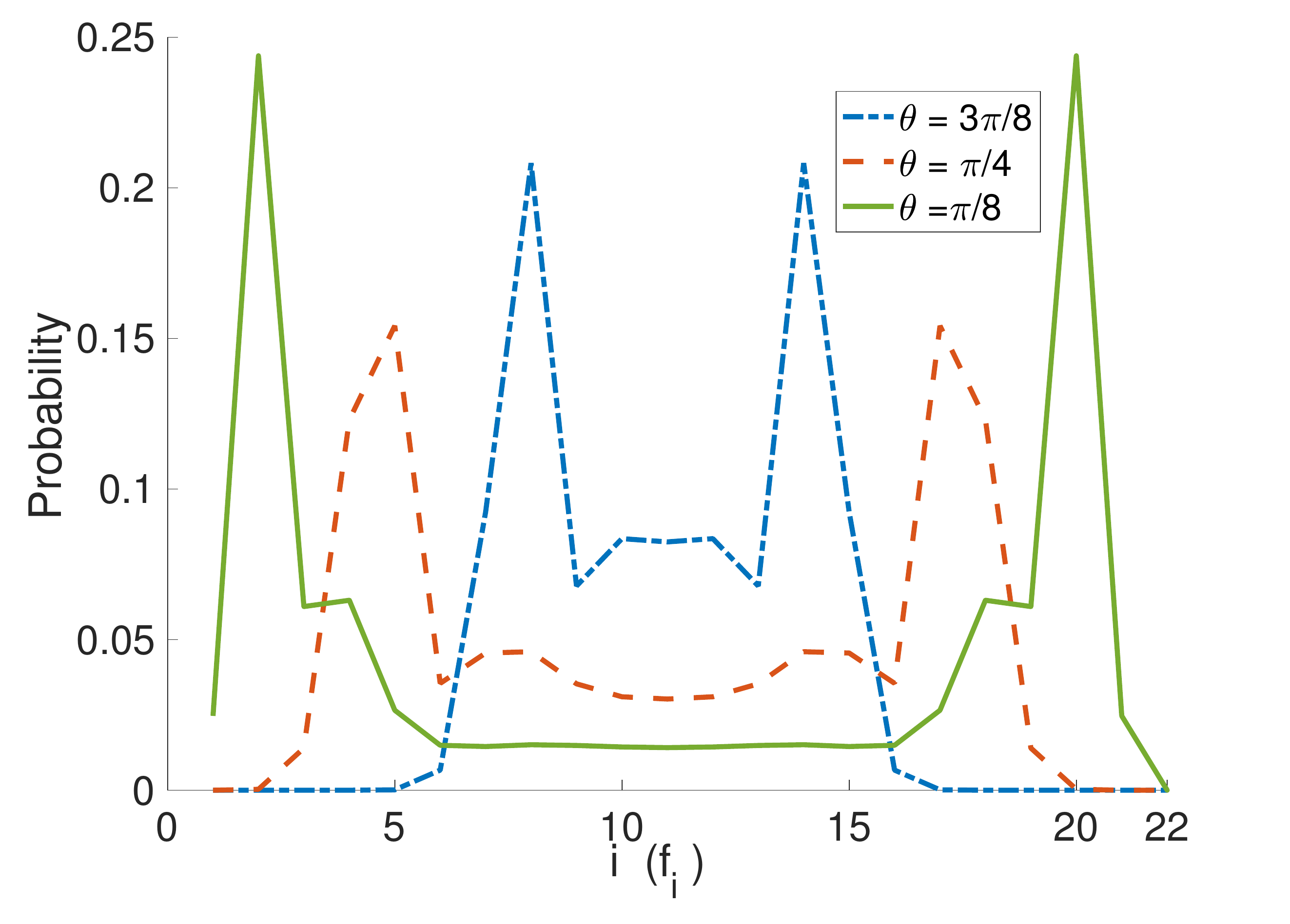}
    \caption{Probability distribution in frequency space after 20 step of walk using different rotation angles $\theta$. Each frequency $f_i$ will be identified with separate position space with non-zero probability in Fig.\,\ref{fig:prob1}.}
    \label{fig:prob2}
\end{figure}

In Figs.\,\ref{fig:prob1} and \ref{fig:prob2} we show the spread position Hilbert space and frequency space after 20 steps of quantum walk when different rotations  in HWPs are used. These distribution can be further controlled by using different rotation angles $\theta$ in each path in combination of $\delta$, initial state parameter. By neglecting the position with zero probabilities in Fig.\,\ref{fig:prob1}, each position labelling can be identified with the frequency.  

We characterize the entanglement between the different degrees of freedom using  negativity,  $\mathcal{N}(\rho) = \frac{\vert\vert\rho^{\Gamma_A}\vert\vert_1-1}{2}$, where $\rho^{\Gamma_A}$ is the partial transpose of $\rho$ with respect to subsystem $A$ and $\vert\vert A \vert\vert$ is the trace norm.
\begin{figure}[h!]
   \centering
    \includegraphics[width=0.47\textwidth]{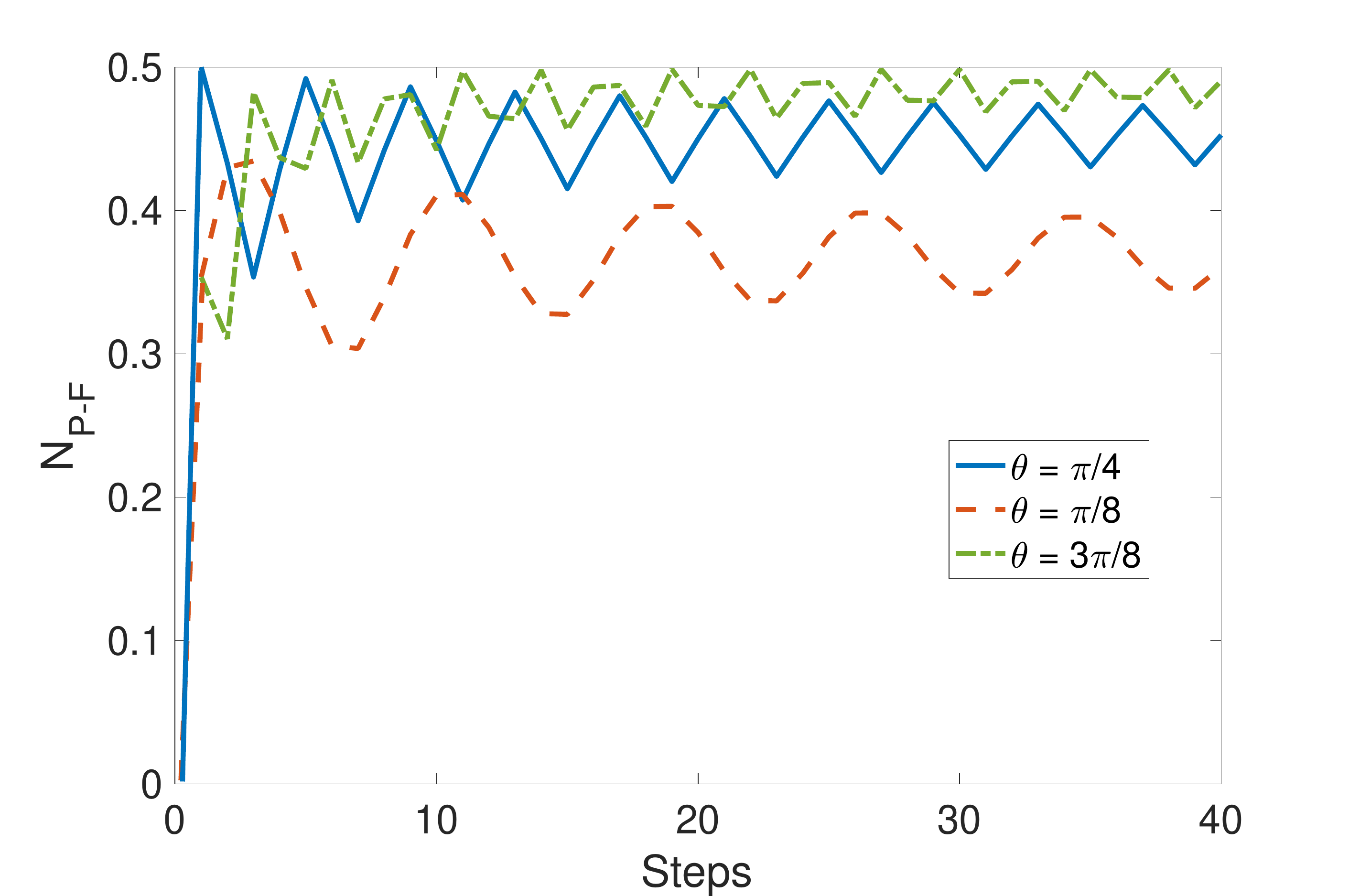}
    \caption{Negativity between polarization and frequency space as function of steps (time).}
   \label{fig:neg1}
\end{figure}
\begin{figure}[h!]
   \centering
    \includegraphics[width=0.47\textwidth]{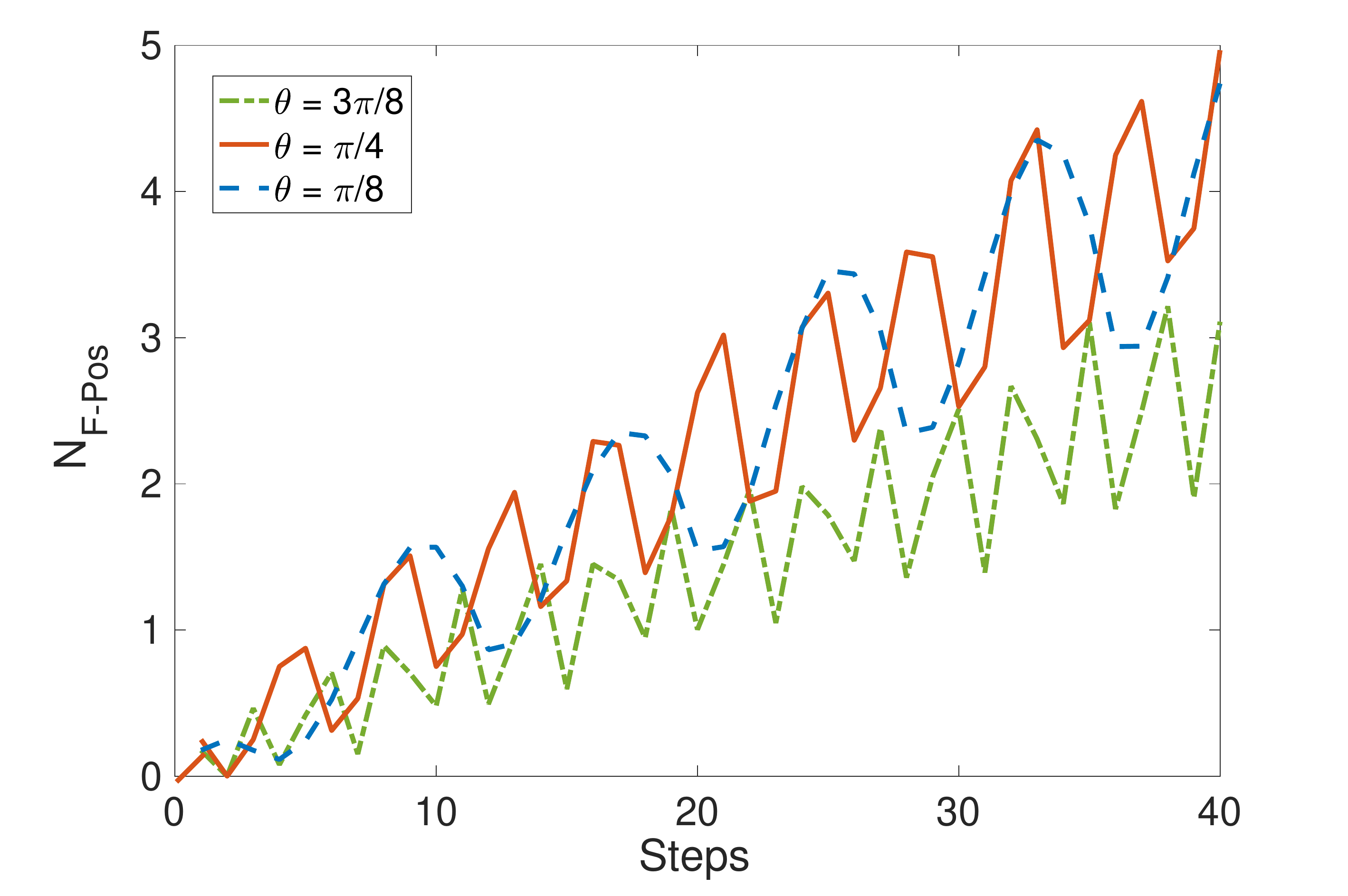}
    \caption{Negativity between position and frequency space as function of steps (time).}
    \label{fig:neg2}
\end{figure}
In Fig.\,\ref{fig:neg1} we show the negativity between polarization and frequency space, $\mathcal{N}_{P-F}$ after taking trace over the position space and a similar behavior can be seen for negativity between polarization and position space $\mathcal{N}_{F-Pos}$. With increase in $\theta$ average value of negativity increases and an inverse is seen for negativity between polarization and frequency space obtained after taking trace over the polarization space. Unlike $\mathcal{N}_{P-F}$ which oscillated around an average value, $\mathcal{N}_{F-Pos}$ increases with number of steps and this is due to the increase in the dimension of both, position and frequency with steps. 

In this work, we have looked at the possibility of manipulating polarization and path degree of freedom to generate an extended dimension of frequency space and generate hyperentangled states. In general resolving closely placed frequencies is an experimental hurdle. For example, a 1550 nm. laser beam has frequency of around 193 THz, while an EOM can produce a frequency shift of the order of 10-15 GHz MHz, which is around $0.005\%$ of the frequency of the laser beam. Traditionally, unwanted symmetric sidebands are invariably produced in EOMs, and we therefore need to filter undesireable components, thereby leading to less efficient frequency shifts. Quadrature and in-phase modulators remove such sidebands using destructive interference among multiple units, while Serrodyne modulators utilise saw-tooth waveforms for the generation of frequency shifts. This, however, is not easily extendible to the GHz regime and requires high-power, broadband electronics. There are also other methods such as spectral shearing \cite{wright2017spectral, fan2016integrated} and adiabatic tuning of optical cavity resonance \cite{preble2007changing}, which can undertake frequency shifts with pulsed operations having a known timing reference. Recently, a quadrature phase-shift keying modulator was used to attain a 15.65 GHz frequency shift\,\cite{chen2021single}. When we have multiple such devices, it get more complicated but in our scheme, the closely placed frequencies are handled by tagging them with position space without need for a special frequency resolver. 

However,  advancements in the field of frequency discriminators and instantaneous frequency measurement subsystems \cite{schilt2011frequency, hartnett2001novel, choondaragh2014microstrip, stec2006multibit, gruchala200211} have been reported. Microstrip-based discriminators have been seen to achieve high sensitivity and high dynamic range, with a recent work being able to attain a frequency resolution of 16 MHz by using open-loop spiral resonators instead of the more conventionally used coupled resonators \cite{rahimpour2018high}. High resolution and accuracy have also been obtained with quadrature microwave frequency discriminators (QMFD) \cite{tsui1990instantaneous, chatterjee2015wide, de20052, lurz2018compact, yingjiao2018research, wang20214}. Adaptive spectrum sensing techniques based on group delay induced by Brillouin scattering in optical fiber has been shown to be able to temporally discriminate two-tone signals with a resolution of 60 MHz \cite{kadum2021slow}. Modern spectrum analyzers can also provide robust and efficient ways to observe and discriminate between frequency modes. Recently, \textit{Pelusi et al} used compact planar rib waveguides based on $\mathrm{As}_2\mathrm{S}_3$ glass, providing a virtual ‘lumped’ high nonlinearity in a monolithic platform that was capable of integrating multiple functions, and output spectral resolution of 1.25 GHz (0.01 nm) \cite{pelusi2009photonic}. Above outlined experimental progress in handling frequency in addition to it control through quantum walk approach shows a huge potential of hyperentangled states in polarization-frequency-position space in quantum information processing applications.

\section{Conclusion}
\label{conc}
 In this work, we present a scheme to pass single photon through an intricate network of concatenated units comprising of a combination of HWPs, PBSs and AOMs and generate hyperentangled state in polarization-path-frequency domain. BY harnessing the power of a polarization-controlled quantum walk one can also tune to the desired frequency from the extended frequency space that can be generated in the system. We have show the entanglement generated between different degree of freedom associated with photon in quantum walk setup. We particularly see a generally rising (normalised) negativity of entanglement between the position and frequency degrees-of-freedom. The addition to generation of hyperentangled states for various quantum information processing tasks, the scheme can also be harnessed to generate closely shifted frequencies and resolve them with tagging with position space.  
 

\textit{Acknowledgement.} MGM would like to acknowledge the insightful discussions with Abhay Hegde and Ameen Yasir. MGM  and CMC acknowledge the support from the Office of Principal Scientific Advisor to Government of India, project no. Prn.SA/QSim/2020.

\bibliography{apssamp}

\end{document}